\documentclass[preprint]{revtex4-1}
\usepackage[pdftex]{graphicx}

\begin{document}

\title{Dynamic jamming of iceberg-choked fjords}

\author{Ivo R. Peters$^1$}
\email[]{irpeters@uchicago.edu}
\author{Jason M. Amundson$^2$}
\author{Ryan Cassotto$^3$}
\author{Mark Fahnestock$^4$}
\author{Kristopher N. Darnell$^5$}
\author{Martin Truffer$^4$}
\author{Wendy W. Zhang$^1$}
\affiliation{
$^1$James Franck Institute \& Department of Physics, The University of Chicago, Chicago, Illinois, USA\\
$^2$Department of Natural Sciences, University of Alaska Southeast, Juneau, Alaska, USA\\
$^3$Department of Earth Sciences, University of New Hampshire, Durham, New Hampshire, USA\\
$^4$Geophysical Institute, University of Alaska Fairbanks, Fairbanks, Alaska, USA\\
$^5$Jackson School of Geosciences \& Institute for Geophysics, University of Texas, Austin, Texas, USA}

\begin{abstract}
We investigate the dynamics of ice m{\'e}lange by analyzing rapid motion recorded by a time-lapse camera and terrestrial radar during several calving events that occurred at Jakobshavn Isbr{\ae}, Greenland.  During calving events (1) the kinetic energy of the ice m{\'e}lange is two orders of magnitude smaller than the total energy released during the events, (2) a jamming front propagates through the ice m{\'e}lange at a rate that is an order of magnitude faster than the motion of individual icebergs, (3) the ice m{\'e}lange undergoes initial compaction followed by slow relaxation and extension, and (4) motion of the ice m{\'e}lange gradually decays before coming to an abrupt halt. These observations indicate that the ice m{\'e}lange experiences widespread jamming during calving events and is always close to being in a jammed state during periods of terminus quiescence. We therefore suspect that local jamming influences longer timescale ice m{\'e}lange dynamics and stress transmission.
\end{abstract}

\maketitle

\section{Introduction}
Several recent studies \cite[e.g.,][]{Joughin2008,Amundson2010,Howat2010,Walter2012,Seale2011,Sundal2013,Foga2014} have suggested that ice m{\'e}lange, or dense packs of icebergs and sea ice found in proglacial fjords, can influence calving of icebergs from tidewater glaciers on seasonal time scales. Variations in glacier length due to seasonal variations in calving can be on the order of several kilometers. Consequently, ice m{\'e}lange may indirectly affect the stability of tidewater glaciers due to the nonlinear relationship between glacier geometry and glacier dynamics \citep{Joughin2012}.

Ice m{\'e}lange forms when ocean currents and surface winds are unable to efficiently evacuate icebergs from a fjord. The persistence of ice m{\'e}lange is a function of iceberg productivity, fjord geometry, and sea ice formation. At some fjords, ice m{\'e}lange exists only when air and water temperatures are low enough to permit the growth of a thick sea ice matrix \citep{Howat2010,Walter2012}. At others, a combination of high iceberg productivity and confining fjord geometry enables ice m{\'e}lange to persist year round as a result of iceberg-iceberg and iceberg-bedrock contact forces \cite[see also][]{Geirsdottir2008,Jakobsson2012}.

Several observations suggest that ice m{\'e}lange can be viewed as a weak, granular ice shelf capable of exerting resistive stresses onto a glacier terminus \citep{Thomas1979,Johnson2004} and influencing iceberg calving. First, seasonal variations in calving rates are well-correlated with the formation and dispersal of ice m{\'e}lange (or changes in mobility) \citep{Sohn1998,Reeh2001,Joughin2008,Howat2010,Seale2011,Walter2012,Cassotto2015}. Second, during periods of terminus quiescence, ice m{\'e}lange is pushed from behind by the advance of the glacier terminus (i.e., the ice m{\'e}lange must also push back against the terminus) \citep{Joughin2008,Amundson2010,Sundal2013,Foga2014}. Third, complete dispersal of ice m{\'e}lange appears to cause a small increase in glacier velocity that is comparable to tidally-induced velocity variations \citep{Walter2012}. Finally, observations and theoretical work suggest that resistive forces from ice m{\'e}lange do not need to be large to hold together heavily fractured termini \citep{Reeh2001,Amundson2010}.

The slow and steady motion of ice m{\'e}lange observed between calving events belies its dynamic and variable behavior. For example, large calving events at Jakobshavn Isbr{\ae} cause icebergs in the fjord to quickly accelerate to speeds of about 1 m/s (an increase in speed by three orders of magnitude). Following calving events the ice m{\'e}lange moves slower than the glacier but gradually accelerates over the subsequent days until it reaches the same speed as the glacier terminus \citep{Amundson2010}. These observations suggest that ice m{\'e}lange undergoes temporal variations in strain that are modulated by terminus activity.

The rheology of ice m{\'e}lange is unknown, and therefore it is not currently possible to calculate the resistance provided by ice m{\'e}lange or fully evaluate its impact on tidewater glacier dynamics. To gain insights into its rheology, we collected high temporal and spatial resolution time-lapse photography and terrestrial radar data at Jakobshavn Isbr{\ae} during a two week period in summer 2012. We focus our analysis on rapid motion of ice m{\'e}lange that was observed during several full-glacier-thickness calving events. 

We analyze our data in light of the recent discovery of dynamic jamming fronts, a phenomenon that occurs in granular systems that are close to the jamming point. A granular system with a packing fraction (i.e., iceberg concentration) below the jamming point can flow, but becomes rigid when the packing fraction reaches or exceeds the jamming point \citep{Cates1998,Liu1998,Reichhardt2014}. Experiments in a variety of systems have shown that the transition to this jammed state can occur as a transient process in which the system jams locally due to a strong perturbation and the jammed region quickly spreads throughout the entire system \citep{Liu2010,vanHecke2012,Waitukaitis2012,Waitukaitis2013,Burton2013,Peters2014}. The details of dynamic jamming, and the types of systems in which it occurs, are only beginning to be explored. Here, we will show that dynamic jamming occurs in closely-packed ice m{\'e}lange, a system that is orders of magnitude larger than other systems in which dynamic jamming has been observed.

\begin{figure*}[t]
\centering
\noindent\includegraphics[width=150mm]{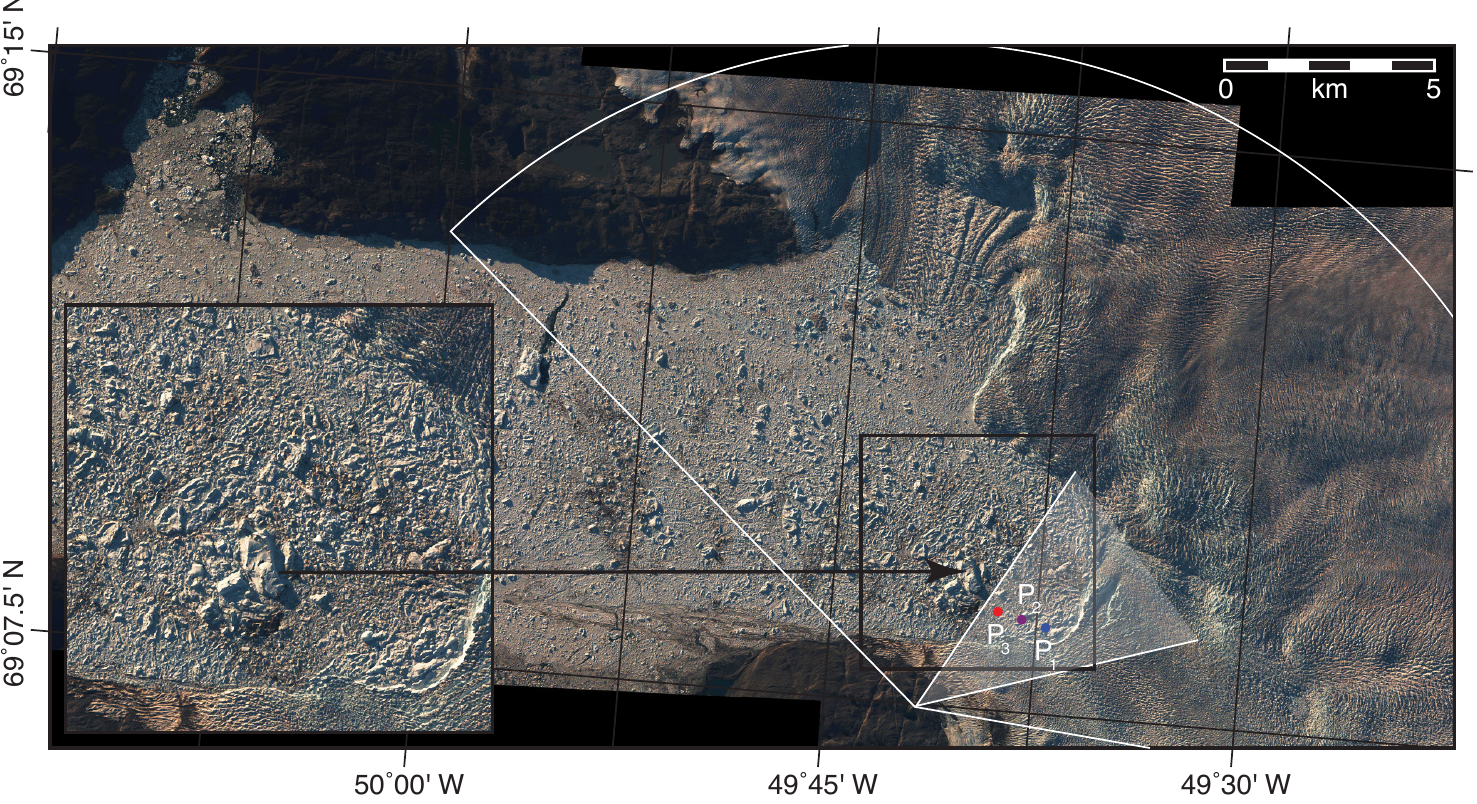}
\caption{WorldView-2 image of Ilulissat Icefjord and the terminus of Jakobshavn Isbr{\ae}, acquired on 6 July 2010 (copyright 2010 Digital Globe, Inc.). The areas covered by the terrestrial radar and the time-lapse camera are indicated by the semi-circular wedge and the shaded triangle, respectively. The inset shows a close-up of the near-terminus region. The blue, purple, and red dots (P1--P3) indicate the approximate geographic coordinates of the pixels that were tracked in the time-lapse photographs and presented in Figure~\ref{fig:kinematics}c.}
\label{fig:map}
\end{figure*}

\section{Methods}
We operated a high-rate time-lapse camera and a terrestrial radar interferometer at Jakobshavn Isbr{\ae} (Figure~\ref{fig:map}) from 30 July--13 August 2012. The instruments recorded the motion of the ice m{\'e}lange and glacier terminus area during several calving events.

The time-lapse camera system consisted of a Canon EOS 40D camera with a 28~mm lens, a Canon Timer Remote Controller (TC-80N3), and a custom-built power supply. The camera was oriented toward the glacier terminus and took one photo every 10~s. The camera clock was set to GPS time and was adjusted every couple of days to correct for clock drift.

A Gamma Remote Sensing GPRI-II radar interferometer \citep{Werner2008,Werner2012,Dixon2012} was used to image the glacier and proglacial fjord. The GPRI-II is a Ku-band ($\lambda=1.74$~cm) radial scanner with a range resolution of 0.75 m and an azimuth resolution of 0.4$^\circ$ (i.e., 7 m at a range of 1 km), and is capable of resolving millimeter-scale deformation via differences in electromagnetic phase. However, rapid motion of the ice m{\'e}lange during calving events results in incoherence, which precludes measurement of the interferometric phase. Therefore, our analysis is based on the radar backscatter signal, also referred to as multi-look intensity (MLI) images. The radar performed 162$^\circ$ scans using a 16-km radius; it scanned every three minutes except for during a few hours of high winds. Radar backscatter data was reprojected to 15-m Cartesian space using slant range. Given the low-grazing angle of the radar in this application, the difference between slant range and horizontal range is less than 1\%.

The time-lapse photography and radar data sets are complementary and together provide a comprehensive view of the ice m{\'e}lange motion during calving events. We analyzed both data sets using a particle image velocimetry (PIV) algorithm to obtain velocity fields. Due to a poor look angle and poor ground control, pixel displacements in the time-lapse photos were not translated into true ground displacements. When processing the time-lapse photos we used a correlation window size of 40$\times$40 pixels. We analyzed the radar data by first using a coarse PIV pass (correlation window size of 64$\times$64 pixels) to be able to detect large displacements, which we subsequently refined in two steps down to 32$\times$32 pixels for increased spatial resolution.

\section{Observations of ice m{\'e}lange motion}
We observed seven full-glacier-thickness calving events. During these events, one to several icebergs with spatial dimensions of hundreds of meters detach from the glacier and subsequently capsize. The detachment and capsize of each iceberg lasts about 5 min; rapid motion of ice m{\'e}lange generally continues for 30--60 min and often terminates abruptly \cite[see also][]{Amundson2008,James2014}.

\subsection{Jamming front and kinetic energy}
\begin{figure}[t]
\centering
\noindent\includegraphics[width=86mm]{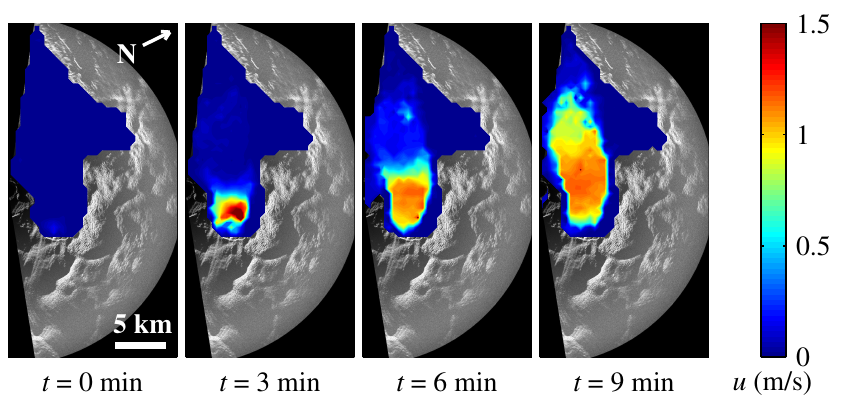}
\caption{Time-evolution of the ice m{\'e}lange velocity field, derived from PIV analysis of radar backscatter data, from a full-glacier-thickness calving event that occurred on 2 August 2012; $t=0$~min corresponds to 23:08:30 UTC.}
\label{fig:PIV}
\end{figure}

The velocity fields derived from time-lapse photography and terrestrial radar data allow us to determine how quickly motion induced by the calving event spreads through the ice m{\'e}lange. Figure~\ref{fig:PIV} shows snapshots of the calculated velocity field during a typical calving event. The calving event induced motion in the innermost fjord concurrent with the detachment and capsize of the calving iceberg. The area of the ice m{\'e}lange that was affected rapidly expanded far down fjord during the following minutes.

\begin{figure}[t]
\centering
\noindent\includegraphics[width=86mm]{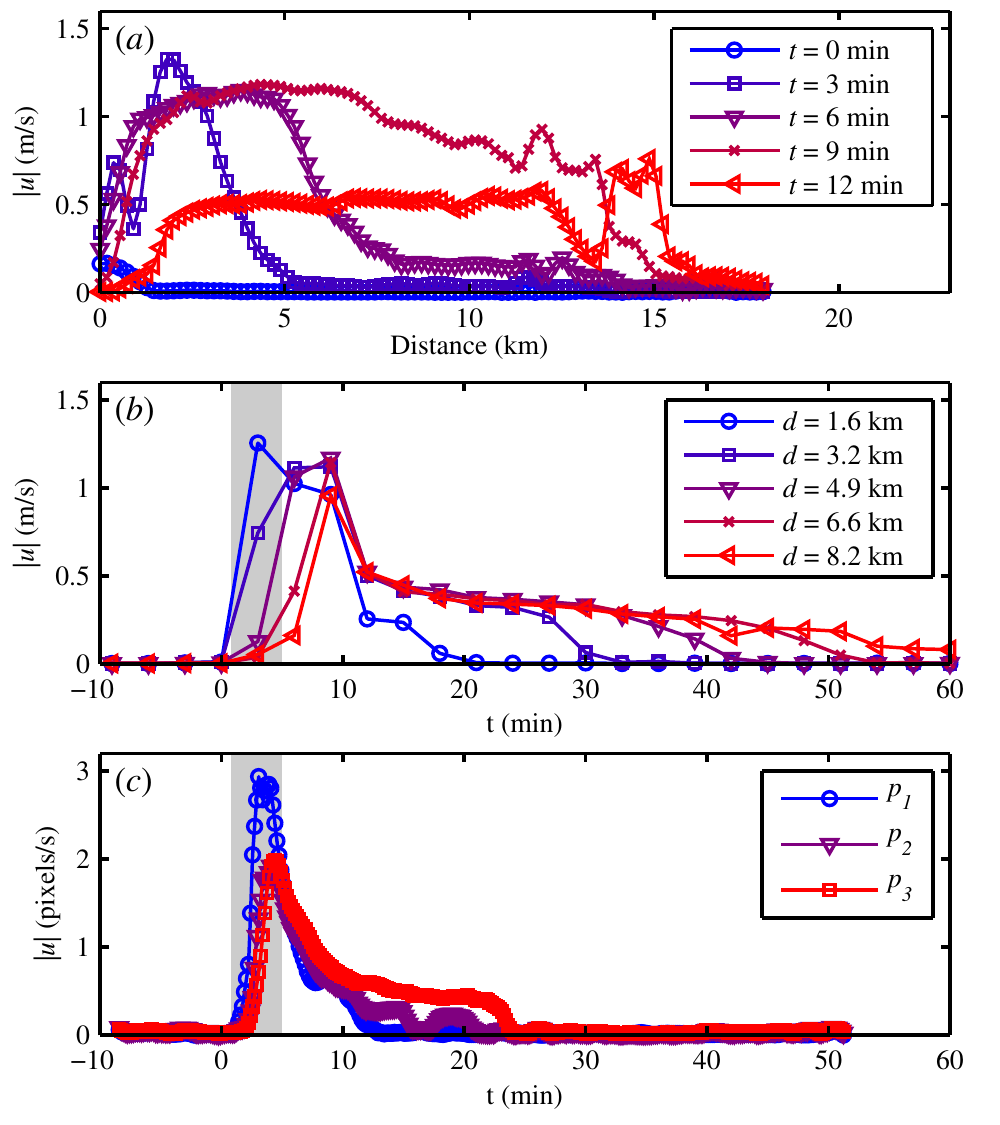}
\caption{Ice m{\'e}lange velocities from the calving event presented in Figure \ref{fig:PIV}. (a) Velocity profiles along the ice m{\'elange} during and after the calving event. The glacier terminus is at $x=0$. (b) Velocity as a function of time at various distances from the terminus. (c) Pixel velocities (from time-lapse photography) at increasing distances from the terminus (from $P_1$ to $P_3$). See also Figure \ref{fig:map}. In (b) and (c), the gray box indicates the period of time during which icebergs are actively calving from the terminus.}
\label{fig:kinematics}
\end{figure}

To quantify the spreading of the motion in the ice m\'elange, we determine the velocity profile along a line from the terminus to the end of the field of view of the radar. Figure~\ref{fig:kinematics}a shows the velocity profile along this transect at 5 instances in time, with 3 minutes between each curve. During the first 9 minutes, a clear front can be observed propagating down fjord; some acceleration of the ice m{\'e}lange occurs prior to the arrival of this front, as can be seen in the purple ($t=6$~min) curve between about 7 to 12 km down fjord. Later, the motion dissipates, resulting in an overall decrease in velocity and the front becoming more difficult to identify. The sudden increase in velocity when the front passes and the subsequent slow relaxation can be seen more clearly by measuring the velocities at specific distances from the terminus and plotting the velocity as a function of time (Figure~\ref{fig:kinematics}b,c). The sudden increase in velocity occurs at a later time at points farther from the glacier. Whereas the acceleration occurs throughout the ice m{\'e}lange within 10 minutes, the relaxation is spread out over about 40 minutes and is slower down fjord. Using a speed threshold of 0.5~m/s, we determine the position of the front in each frame, from which we determine a typical speed at which the front propagates. We calculated front propagation speeds in the range of 16 to 20~m/s, an order of magnitude faster than the typical speeds of individual icebergs. Variations in front speeds may reflect variations in packing fraction or forcing mechanism (see Section \ref{sec:Interpretation}).

We also estimate the amount of kinetic energy, $K$, as a function of time in the ice m{\'e}lange by evaluating $K=\frac{1}{2}\rho_i h \int\int(u_x^2+u_y^2)dxdy$, where $\rho_i=917$~$\rm{kg/m^3}$ is the density of ice, $h\sim100$~m is the thickness of the ice m\'elange (iceberg freeboard ranges from about 1 m to greater than 10 m), and ${\bf u}=\langle{u_x,u_y\rangle}$ is the velocity field. We neglect the motion of water in our calculations. Peak kinetic energies are on the order of $10^{12}$~J, which is 2 orders of magnitude smaller than the amount of energy that is released by the calving events \citep{MacAyeal2011,Burton2012}.

\subsection{Strain}
We employed a Voronoi analysis \citep{Aurenhammer2013} to investigate whether the ice m{\'e}lange experiences extension or compression during the calving event because we found that the velocity fields (Figure~\ref{fig:PIV}) were too noisy to be used for calculating spatial derivatives. To generate Voronoi diagrams we developed a tracking algorithm, based on cross-correlations, that allowed us to select several large and easily identifiable icebergs (reference points) in one radar image and track it through the subsequent images. To improve performance, the algorithm used the PIV data (Figure~\ref{fig:PIV}) as an initial guess for the displacement to limit the searching area. The Voronoi cells were then computed by determining the proximity of the selected icebergs to each pixel in the image (see inset in Figure~\ref{fig:voronoi}).

During the propagation of the jamming front there is an overall compaction that occurs in about 10 minutes. The fast compaction is followed by a slow relaxation process with a typical timescale of one hour. The packing fraction of the tracked region typically decreases after the relaxation phase. The initial compaction represents the total compaction of the entire tracked region, including a portion of the ice m{\'e}lange to the north that does not experience any motion during the calving event, and therefore underestimates local compaction that occurs as the jamming front passes through the ice m{\'e}lange. Noise in the data prevents us from calculating local compaction. We therefore only use the fractional changes in the areas of the Voronoi cells for order-of-magnitude comparisons with theoretical work on jamming. The total compaction and relaxation and the timescales of ice m{\'e}lange motion differed between calving events, but all events exhibited the same qualitative behavior.

\section{Interpretation}\label{sec:Interpretation}
\begin{figure}[b]
\centering
\noindent\includegraphics[width=86mm]{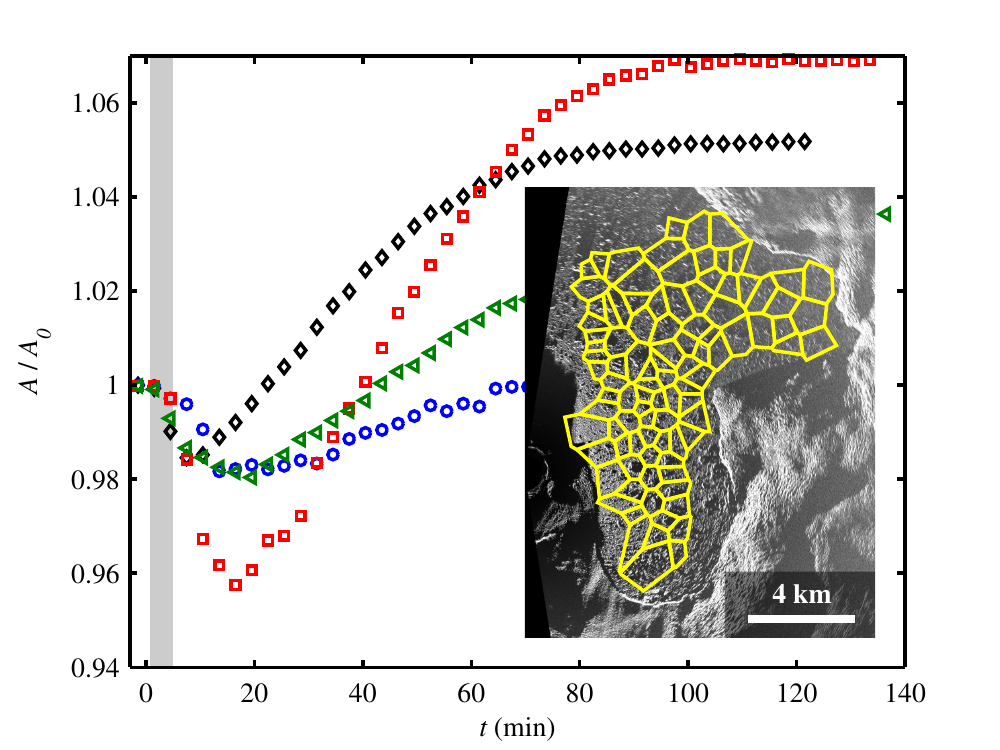}
\caption{Voronoi-cell analysis reveals that the total surface area of the ice m{\'e}lange decreases immediately following calving events (at $t\approx 0$) and then slowly expands. Compaction and relaxation are shown for four calving events. The black diamonds and gray box correspond to the calving event presented in Figures \ref{fig:PIV} and \ref{fig:kinematics}; the inset shows the distribution of the Voronoi cells for that event at $t=0$.}
\label{fig:voronoi}
\end{figure}

We have observed large calving events causing the rapid growth and down fjord propagation of a compacted region of ice m{\'e}lange that moves at a nearly uniform speed. Similar behavior has been observed in other types of dense particulate systems subjected to compressive forcings, such as two-dimensional packing of frictional disks \citep{Waitukaitis2013} and temporary solidification of a layer of dense corn starch suspension floating on oil \citep{Peters2014}. Despite great differences in characteristic scales and composition, these systems are comprised of fairly rigid constituents that strongly resist compression when jammed together but offer negligible resistance when there is finite interstitial spacing between them. Strong compressive stresses cause an initial reduction in the interstitial spacings, but eventually the constituents jam together and prevent further compression. Instead the particulate systems respond by propagating the jammed region over an increasingly larger area.

The similarity between these systems suggests the following interpretation of the observed ice m{\'e}lange motion. Between calving events the ice m{\'e}lange is closely-packed but is not at the jamming point everywhere (i.e., the fjord is not uniformly jammed). Jamming may still occur locally and have important consequences for stress transmission and ice m{\'e}lange motion. When an iceberg calves, it pushes the icebergs in the inner fjord and compresses the ice m{\'e}lange locally, causing the local packing fraction to rapidly increase until it reaches the jamming point. As a result, there is a large area of ice m{\'e}lange that moves as a single solid mass, which then compresses the ice m\'elange farther down fjord and causes the area of rapid motion to expand. This process continues as long as there is a driving force from the calving iceberg or enough inertia in the jammed region, and can be observed as a jamming front traveling down the fjord at a speed much larger than the speed of the calving iceberg. The jamming front is defined as the region that delineates the jammed regions from the unjammed region down fjord. Friction and drag eventually cause the ice m{\'e}lange to gradually decelerate, but only after some relaxation and expansion has occurred.

The dynamic jamming scenario allows for a quantitative prediction for how quickly the jammed region should grow in response to a compressive forcing. Under compression, the spacing between icebergs reduces because water can freely flow out of the interstitial spacings and because icebergs may be able to partially raft over each other and/or undergo rotations. After an initial compaction the mechanical resistance greatly increases and the icebergs become jammed. A reasonable first approximation of the ice m{\'e}lange motion is that only the area occupied by the water contracts under compression and that the map-view area occupied by icebergs is conserved. Thus, if $\phi$ is the two-dimensional solid packing fraction, then conservation of iceberg area requires
\begin{equation}
\frac{\partial{\phi}}{\partial{t}} + \nabla\cdot(\phi {\bf u}) = 0. 
\label{eq:conservation}
\end{equation}
Below the jamming point, $\phi_J$, the ice m{\'e}lange contracts while offering only a small amount of resistance. Once $\phi$ reaches $\phi_J$, the resistance to any further significant compression jumps discontinuously and becomes far larger than the typical forcing. Instead, continued forcing creates a jammed region at $\phi_J$ that grows in spatial extent.

To estimate the speed of the jamming front, $u_f$, we make the following simplifying assumptions: (i) the velocity field within the jammed region of the ice m{\'e}lange is a spatially uniform one-dimensional translation with icebergs moving at speed $u_J$, (ii) the entire jammed region is at $\phi_J$, (iii) down fjord from the jammed region, the ice m{\'e}lange is barely disturbed and therefore reasonably approximated as a region of zero flow at a uniform initial packing fraction $\phi_0$.  The measured velocity profiles within the ice m{\'e}lange (Figure \ref{fig:kinematics}a) are consistent with these simplifications.  

Conservation of iceberg area then simplifies to a jump condition across the moving front of the jammed region. This condition is most simply enforced by switching to a comoving reference frame in which the jamming front appears stationary: icebergs ahead of the jamming front have speed $-u_f$, icebergs behind the jamming front have speed $-u_f+u_J$, and $\partial \phi/\partial t=0$. Thus, integrating Equation \ref{eq:conservation} across the jamming front yields the jump condition
\begin{equation}
-\phi_0 u_f - \phi_J (- u_f + u_J)=0,
\end{equation}
which rearranges to give 
\begin{equation}
u_f=u_J\left(\frac{\phi_J}{\phi_J-\phi_0}\right).
\label{eq:front_speed1}
\end{equation} 

Due to the chaotic nature of ice m{\'e}lange, we can not readily calculate packing fractions from our images. In two-dimensions, the packing fraction is defined as
\begin{equation}
\phi=\frac{A_{\rm ice}}{A},
\label{eq:phi}
\end{equation}
where $A_{\rm ice}$ is the effective area occupied by ice within a control area $A$. The control area can change dynamically due to compression, but the area occupied by icebergs remains constant under our first order approximation (Equation \ref{eq:conservation}). Plugging Equation (\ref{eq:phi}) into Equation (\ref{eq:front_speed1}) and rearranging gives
\begin{equation}
u_f=u_J \left( \frac{1}{1-\frac{A_J}{A_0}} \right),
\label{eq:u_f}
\end{equation}
where $A_0$ is the initial control area and $A_J$ is the control area after becoming jammed. Equation (\ref{eq:u_f}) indicates that the speed of the jamming front is determined by the proximity of the initial packing fraction to the jamming point. From our observed front speeds of 16--20~m/s and typical horizontal speeds of individual icebergs of the order of 1~m/s (Figure~\ref{fig:PIV}, see also \citep[]{Amundson2010}), we estimate the numerical value of $A_J/A_0$ to be 0.94--0.95. This corresponds to compression of 5--6\%, which is in agreement with the order of magnitude compression shown in Figure~\ref{fig:voronoi} and therefore gives further evidence that dynamic jamming occurs in ice m{\'e}lange during large-scale calving events.

\section{Conclusions}
Analysis of time-lapse photography and terrestrial scanning radar data from Jakobshavn Isbr{\ae} indicates that large portions of proglacial ice m{\'e}lange compact and become uniformly jammed during calving events. A jamming front, which delineates jammed and unjammed portions of the fjord, propagates down fjord at about 20 m/s, or roughly an order of magnitude faster than individual icebergs. After calving activity ceases the ice m{\'e}lange slowly relaxes and decelerates before coming to an abrupt halt. The initial compaction and rapid jamming front are consistent with laboratory studies of jamming in a variety of systems, which together indicate that the ice m{\'e}lange must be close to the jamming point prior to the initiation of the calving events. Thus, local jamming and unjamming, which will likely be difficult to observe in the field, may contribute significantly to the mechanics of ice m{\'e}lange during periods of terminus quiescence. Our observations are from summer, when ice m{\'e}lange has relatively high mobility; ice m{\'e}lange is likely closer to being in a jammed state during winter, which would result in faster jamming fronts during calving events (if and when they occur). Two features of our data, the moderate acceleration of icebergs that precedes arrival of the jamming front and the abrupt halt in motion at the end of the event, have not been observed in other systems. Neither of these observations can be explained by a simple theory of dynamic jamming and therefore require additional study.

Ice m{\'e}lange is easily the largest system that has been observed to experience dynamic jamming. In fact, its difficult to imagine other granular systems on Earth that contain larger constituents. Our study of jamming on the meter-to-kilometer scale is complementary to current efforts in condensed matter physics to extend jamming from macroscopic particulate systems down to the micron-scale and molecular systems (e.g., colloids and block-copolymer lipid system in langmuir troughs). We expect the extrapolation to larger length scales will reveal new regimes of jamming dynamics not accessible in systems with smaller constituents. Furthermore, given the potential impact of ice m{\'e}lange on iceberg calving and glacier stability, this work motivates further investigation of jamming and stress transmission through ice m{\'e}lange and other quasi-two dimensional materials.

\begin{acknowledgments}
Funding for this project was provided by the US National Science Foundation (ANT0944193 and DMR0820054), the Gordon and Betty Moore Foundation (GBMF2626 and GBMF2627), and NASA's Cryospheric Sciences program (NNX08AN74G). Please contact the corresponding author for the data used to generate the figures. We thank Heinrich M. Jaeger, Douglas R. MacAyeal and Sidney R. Nagel for helpful discussions.
\end{acknowledgments}

\end{document}